# Decompaction-wave propagation in a vibrated fine powder bed

Prasad Sonar and Hiroaki Katsuragi
*Department of Earth and Space Science, Osaka University, 560-0043, Japan.*

(Dated: July 18, 2022)

We experimentally study the crack formation and decompaction-wave propagating in a vibrated powder bed consisting of glass beads of 5 $\mu$m in diameter. The vibrated powder bed exhibits three distinct phases depending on the vibration conditions: consolidation (CS), static fracture (SF), and dynamic fracture (DF). Particularly, we found an upward wave propagation in the DF regime when the powder bed is strongly vibrated. As a remarkable feature, we found that in fine cohesive powders, the decompaction-wave propagation speed normalized to gravitational speed is independent of the shaking strength. This result implies that the wave propagation speed is governed by the balance between gravity and cohesion effect rather than vibration strength. We also explore the universality of wave propagation phenomenon in coarser and low-density granular powders.

## I. INTRODUCTION

In industries like food, pharmaceutical, chemicals, and cosmetics, granular powders are the most widely used fundamental material. Various standard processes such as segregation, mixing, compaction, and agglomeration need to be performed via basic activities like storage, transportation and delivery. The physical characteristics of powders are very sensitive to the processing/operating conditions, such as ambient conditions, interaction with structures, prior processes and duration of those processes. Nevertheless, its physical understanding is still in its infancy mainly due to the complex nature of powder materials.

Fluidization of the powder bed is one of the most important processes in powder technology. In past, industrial researchers defined and studied granular powders [1–3], and also suggested various ways to fluidize the powders: gas fluidization [1, 4–7], using finer-particles as additives [8, 9], and applying vibrations to the gas-fluidized bed [4, 10, 11]. To characterize flowability, Geldart classified powders into several types, based on the grain size and density [1]: group A (easily fluidized), B (fluidized readily, causes bubbling, small bed expansion), C (fine, very cohesive particles, lifts as a plug and forms channels), and D (coarse grains, forms stable spouted beds). In usual granular physics studies, coarse grains categorized into group D have been used as typical granular materials. In this study, however, we focus on the dynamics of group C powder which is difficult to be fluidized.

When a granular bed is subjected to the weak or strong mechanical vibration/tapping, compaction or dilation are caused, respectively (e.g., [12–17]). Under strong vibration conditions, non-cohesive (group D) granular bed can be fluidized, in general. Mechanical vibration can also be used to control the flowing state of various powders [4, 11]. Certainly, the effect of external vibrations on the physical characteristics and dynamic behavior of powders is of major industrial interest [18–20]. However, physical understanding of the vibrated group C powder (without gas fluidization) is far from sufficient. In this study, we focus on the effect of mechanical vibration on dynamics of fine cohesive powder.

Fluidization process of coarse (group D) granular matter was systematically investigated [16]. The authors have observed various phenomena depending on vibration strength. Two dimensionless numbers characterizing the vibration strength have been used. One is non-dimensional acceleration $\Gamma = a\omega^2/g$, where $a$, $\omega$, and $g$ are vibrational amplitude, angular frequency, and gravitational acceleration, respectively. The other is shaking strength, $S = a^2\omega^2/gd$, where $d$ is grain diameter. The former is appropriate to characterize relatively weak vibration and the latter is better suited as a scaling parameter in a strongly vibrated state. According to [16], in a vibrated coarse (group D) granular bed, one can observe liquid-like and gas-like phases by increasing vibration strength. Pak *et al.* [21], in their vibrated powder (group A) experiments, observed a layer of gas/air bubbles forming at the shaker base when a critical $\Gamma$ is reached. The mean size of the bubbles was observed to be increasing with $\Gamma$. However, the rate of bubbling was found to be not very sensitive to $\Gamma$. Above the critical value of $\Gamma$, convective rolls were observed with flow from the bottom towards the top.

The purpose of this study is to evaluate phase variations in the vibrated fine cohesive (group C) powder. As a consequence, we found interesting behaviors in the vibrated fine cohesive powder bed, brittle crack formation and propagation of decompaction-waves. In addition, it turned out that the decompaction-wave propagation speed is independent of vibration strength $S$. In this paper, details about these findings are reported.

The organization of this paper is as follows. We present the details of the experiments in §II. The behavior of vibrated powders and the corresponding changes in phases are observed in §III. §IV discusses the decompaction-wave velocity and its relation with the vibrational parameters. In the same section we discuss the cohesive strength estimates. Finally, we conclude in §V. The additional details such as experimental condition dependencies are included in Appendices for more comprehensive understanding.

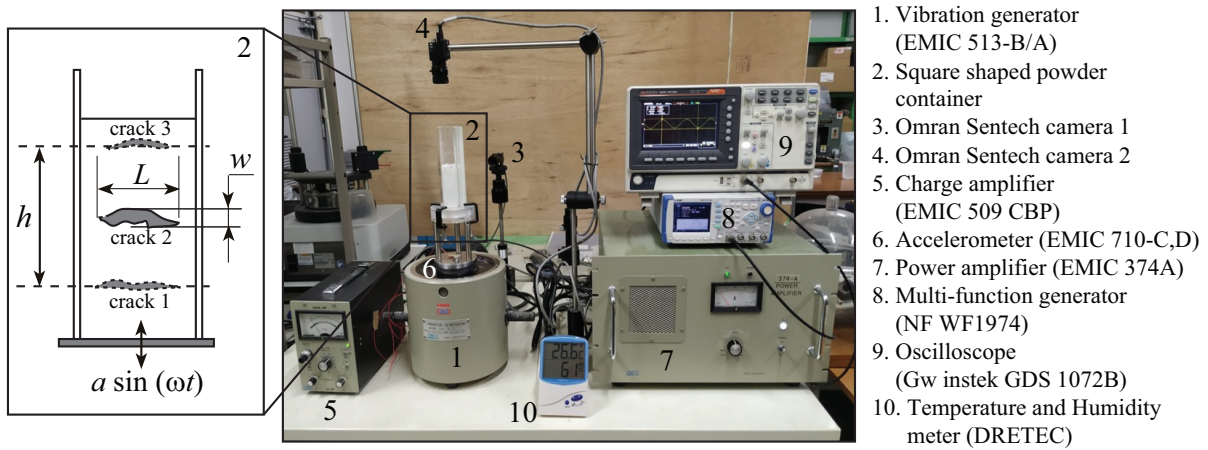

FIG. 1. The experimental setup of the vibrated granular powder bed. The inset shows a schematic of the vibrated fractured powdered bulk consisting of horizontal cracks. In DF phase, these cracks would open in an successive manner from bottom to the top giving an impression of vertically upward traveling wave. The parameters $w$, $L$ and $h$ are crack width, crack length and a traveling height of the crack wave, respectively. We use ImageJ [22] to obtain the crack measurements.

## II. EXPERIMENTS

Figure 1 shows the experimental setup. The square shaped container of inner dimensions $28\,\text{mm} \times 28\,\text{mm} \times 150\,\text{mm}$ is mounted on the vibration generator. The container is made of acrylic (PMMA) with flat and smooth sidewalls. The vibration frequency is set on the multi-function generator, and amplitude is controlled by the power amplifier. The vibration amplitude and acceleration can be measured using an accelerometer. We use glass beads of typical diameter $5\,\mu\text{m}$ (Potters Ballotini, EMB-10, size range is 2–10 $\mu$m and true density $\rho = 2600\,\text{kg/m}^3$) for a model powder material. According to Geldart's chart, such fine powder is classified as Group C powder [1]. Due to the fine grain size, this powder shows cohesive characteristics forming agglomerates in which the inter-particle cohesive force dominates over gravitational force [23]. In the experiment, we first poured the powder into the container. Then, the entire powder bed in the container was vertically vibrated. The mass of powder in the container was varied from 5 g to 50 g. These masses correspond to the initial height of 11 mm and 97 mm, respectively. The initial condition for all experiments was maintained by setting the identical powder volume fraction, $\nu \approx 0.25$. The frequency of vibration was then set prior performing experiments over varying amplitudes. The ranges of amplitude and frequency $f = \omega/2\pi$ are 0.3 mm to 3.6 mm and 25 Hz to 70 Hz, respectively. The videos are recorded using cameras that capture images at 90 fps with spatial resolution 0.058 mm/pixel and $2048 \times 2048$ image size in the presence of fluorescent light (room light). We performed experiments under two different ambient conditions with relative humidity (RH) at 40 % and 85 %.

## III. OBSERVATIONS AND RESULTS

Figure 2(a) shows the phase diagram of the observed phenomena when the powder is vibrated at $f = 30\,\text{Hz}$ and RH = 40 %. As shown, we identify three distinct phases of powder behavior (consolidation, static fracturing, and dynamic fracturing) when different powder masses $m$ (x-axis) are vibrated with various amplitudes $a$ (y-axis).

The region denoted by horizontal lines in the phase diagram represents the consolidated (CS) regime. The corresponding actual images are shown in Fig. 2(b) & (c). The corresponding movies can be found in Supplementary Material (SM). In this regime, we observed condensation of powder with a particular feature of non-deforming stable voids. We observed the powder getting consolidated at low amplitudes ($a \leq 1.5\,\text{mm}$), irrespective of the mass in the container. More importantly, small powder bed ($m < 15\,\text{g}$) is always consolidated even in the strongly vibrated regime ($a > 3\,\text{mm}$). In the previous studies, such a consolidation phase could only be observed under the weak vibration conditions [13, 15, 20, 25, 26]. However, here we found that fine powders are consolidated even when the vibration amplitude is much larger than the grain size. Cohesive effect of the powder presumably sustains the consolidated state. On the contrary, for coarse granular materials, liquid or gaseous states were typically observed when strongly vibrated/energized. Specifically, the phenomena like convection, Leidenfrost, and undulations were observed [16].

When the vibration is further strengthened, different phases can be observed. When both the mass and vibration amplitude exceed certain thresholds ($m > 15\,\text{g}$ and $a > 1.5\,\text{mm}$), the existing voids start deforming into the horizontal cracks in the consolidated bulk that indicate local fractures in the consolidated powder. This means that brittle fracturing can be observed in a strongly vibrated fine powder whereas only the fluidization (i.e., melting) is observed in a strongly vibrated coarse granular matter. This fracturing regime is named static fracture (SF) phase (shown by red colored vertical dashed lines in Fig. 2(a)), which can be easily differentiated from the CS regime based on qualitative observations. See Appendix B for the details. An actual image of this phase is shown in Fig. 2(d). As shown, the dark horizontal lines representing cracks span through the entire container width, and thus, also assumed to extend through the depth of container as well. The critical vibration strength at the boundary between CS and SF phases corresponds to the minimum energy causing cracks that break the cohesive bonds of the fine powder particles. Besides, the vibration strength must overcome the gravitational effect as well. The cracks generated in the bulk are layered horizontally as shown in Fig. 2(d). Similar structure has been observed earlier in a gas-fluidized powder bed with vibration [11]. The local fractures occurring at the top free surface allow the expansion of the powder bed, and thus, also result in the formation of small sized agglomerates (see Appendix C for more details). Throughout the experiments, we observed continuous formation and fragmentation of agglomerates largely on the top surface.

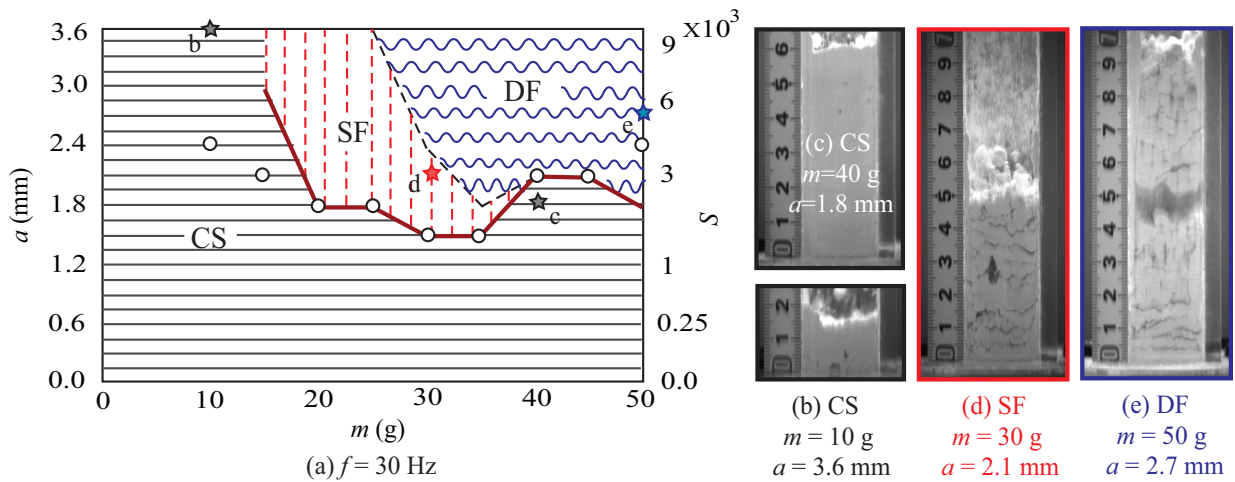

FIG. 2. (a) Phase diagram for $f = 30\,\text{Hz}$, $\text{RH} = 40\,\%$, and the corresponding raw images of: (b) and (c) consolidated state (CS), (d) static fracture (SF), and (e) dynamic fracture (DF). The real time videos of three phases can be found in Supplemental Material (SM) [24]. The red line and hollow circles on the phase diagram indicate the boundary separating consolidated state and the fractured state for $\text{RH} = 40\,\%$ and $\text{RH} = 85\,\%$, respectively. The star pointers on the phase diagram indicate the data points for the corresponding raw images. The non-deforming stable voids in CS regime can be qualitatively differentiated with SF regime, where stable voids deform into fine cracks and ruptured top surface. DF regime is quantitatively differentiated using criterion based on crack area, which should be greater than $10\,\text{mm}^2$. Details of phase classification is written in Appendix B.

Further increase in vibration strength for increased bed mass caused an intriguing phenomenon: upward propagation of the horizontal crack [24]. We define this regime as dynamic fracture (DF) phase (shown by blue colored horizontal wavy lines in Fig. 2(a) and an image in Fig. 2(e)). In this regime, decompaction of the powder bed by the crack opening propagates vertically. (see the corresponding video in the SM [24]). Actually, this decompaction-wave propagation consists of the repeated cycles of successive crack openings from bottom to top. To the best of our knowledge, this type of decompaction-wave propagation in the vibrated powder bed has not been reported in literature.

The hollow circles in Fig. 2(a) indicate the fracturing boundary between CS and SF (or DF) when the experiments are performed at $\text{RH} = 85\,\%$. The overlap of two boundaries, for $\text{RH} = 40\,\%$ and $85\,\%$, suggests the negligible effect of RH on the vibrated powder behavior. In general, the relative humidity plays an important role on the powder behavior. The most of group A powders can behave like group C powders when fluidized with air of RH in the range $60\,\% - 90\,\%$ [27]. The powder behavior changes when aerated with humid air. However, in our experiments, relative humidity of the laboratory has changed within typical range of $40\,\% - 85\,\%$. Moreover, we use no air for fluidization, and the powder used for experiments is carefully handled. We performed experiments by following the exact procedure as mentioned earlier under two different humidity conditions of $40\,\%$ or $85\,\%$, respectively. Figure 3 shows the phase diagram for $\text{RH} = 85\,\%$.

Figure 4 shows the phase diagrams for frequencies $25\,\text{Hz}$ and $40\,\text{Hz}$, respectively. Basically, global structure of the phase diagrams are qualitatively similar. However, we can also confirm the differences. At $f = 25\,\text{Hz}$, we observed that consolidated state dominates. And it requires much higher amplitude to cause the fracture in consolidated (CS state) bulk. On the contrary, at $f = 40\,\text{Hz}$, the fracturing occurs at low amplitude values.

For the quantitative characterization of the DF phase, we measured the decompaction-wave propagation speed.

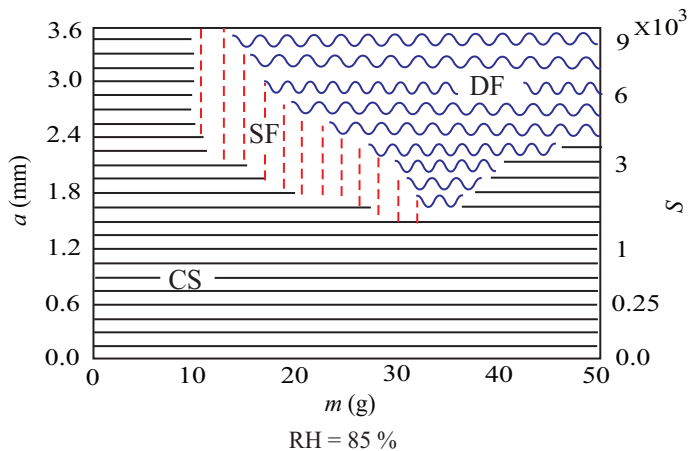

FIG. 3. The Phase diagram for $f = 30\,\text{Hz}$ and RH = 85%.

TABLE I. Comparison of $K$ for various experimental conditions.

| Experimental conditions | K |
| --- | --- |
| Square container (RH = 40 %) | $0.1600 \pm 0.04$ |
| Square container (RH = 85 %) | $0.1428 \pm 0.06$ |
| Circular container (RH = 40 %) | $0.1760 \pm 0.03$ |

The ImageJ [22] was used to analyze the experimentally obtained images. The schematic presented in the left of Fig. 1 shows the void/crack parameters that we measure in the image analysis. Figure 5 shows an example of the decompaction-wave propagation. In the analysis, the largest crack in each frame was followed and the other minor cracks were ignored and filtered out. This procedure was applied for all frames as long as the significantly large crack whose area is greater than 10 mm² is visible. This is also a quantitative criterion we set to distinguish DF regime from others. Then, by measuring the length traveled by crack $h$ and the actual time difference between the initial and final frames $\Delta t$, we calculate the average decompaction-wave propagation speed $v = h/\Delta t$. We take an average of 3 to 5 different events of decompaction-wave propagation for every parameter set. The measured average values of crack width $w$, obtained as the standard deviation about the mean position of the crack, length $L$, and traveled height $h$ for the case of $m = 50\,\text{g}$, $f = 30\,\text{Hz}$, and $a = 2.4\,\text{mm}$ (Fig. 5) are $3.66 \pm 0.25\,\text{mm}$, $23.27 \pm 0.32\,\text{mm}$, and $49.41 \pm 4.84\,\text{mm}$, respectively. By applying this method to all DF data, parameter dependence of the decompaction-wave propagation speed can be evaluated.

## IV. ANALYSIS AND DISCUSSION

To discuss the relation between decompaction-wave propagation speed and vibration strength, they should be properly normalized. Here, we employ a normalized characteristic speed $v/\sqrt{gw}$ and shaking strength $S = a^2\omega^2/gd$ to characterize the dimensionless relation between propagation speed and vibration condition, respectively.

### A. Role of shaking strength

In Fig. 6, we plot $v/\sqrt{gw}$ vs. $S$ with various frequencies, amplitudes and masses of powders at RH = 40 % in DF regime. As displayed, it is difficult to confirm the clear correlation between $v/\sqrt{gw}$ and $S$. Rather, $v/\sqrt{gw}$ values simply scatter around a certain average value $K = 0.16$, namely,

$$\frac{v}{\sqrt{gw}} = KS^0. \quad (1)$$

This behavior is in contrast to the motion observed in a vibrated coarse granular system. The vibration-induced granular convective speed is scaled by $S$ (or equivalently the vibration speed) [28, 29]. In this experiment, however, the $S$-independent $v/\sqrt{gw}$ behavior can be confirmed in all experimental conditions. As mentioned earlier, the humidity condition has negligible effect on the powder behavior. We find $K = 0.16 \pm 0.04$ and $0.142 \pm 0.06$ at RH = 40 % and 85 %, respectively. We confirm this $S$-independent behavior by performing specific experiments using circular shaped container of similar base area with inner diameter of 32 mm. We find marginal increase in $K$. Table I compares $K$ for three different experimental conditions. In this analysis, we employ $S$ to characterize the vibration strength because the vibration amplitude is much greater than the grain size. Even if we use $\Gamma$, any clear correlation cannot be confirmed (see Appendix D for details).

In [30, 31], similar crack formation/propagation in decompaction process of a free-falling granular column was reported. In their system, the successive cracks preferentially occurring at the lower part split the pile into several blocks and cause an ascending decompaction-wave in the bulk. Moreover, the equivalence among the decompaction process of falling down and the upward motion of a pile in a continuously vibrated box was also mentioned [19]. However, the material used in their study was non-cohesive coarse aluminium beads ($d = 1.5\,\text{mm}$, group D). Further, this phenomenon was related to the surface roughness of the lateral boundaries. The cracks originated by the side walls are oriented to oppose to the motion of the material relative to the lateral walls. This is not the case in the current experiment of the vibrated powder. We observe roughly horizontal cracks which are opened around the central region and ascend. In the vibrated powder bed, cohesive effect must be considered to explain the observed phenomena.

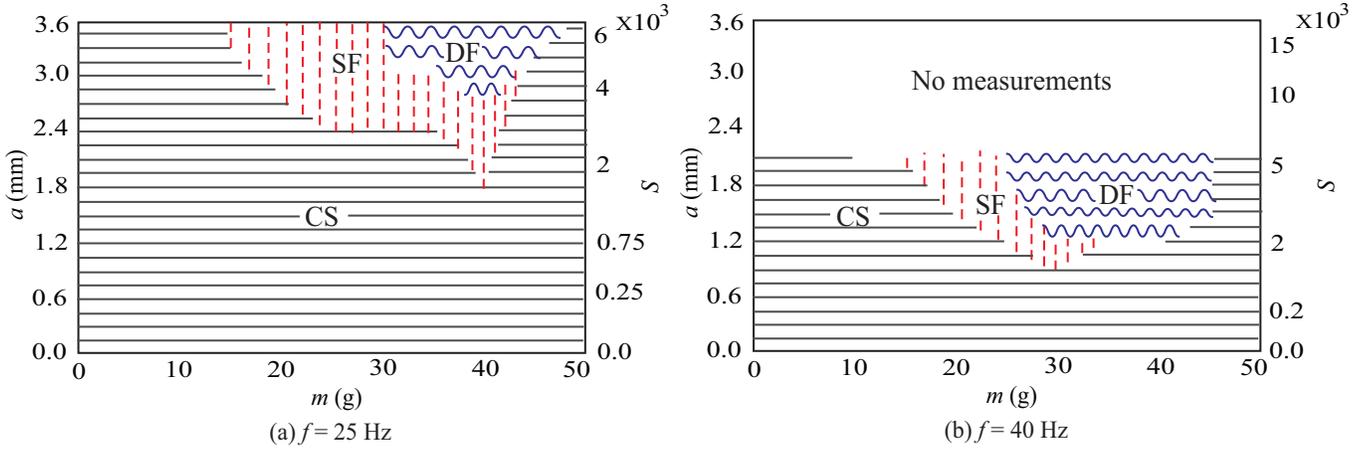

FIG. 4. The Phase diagrams at RH = 40 % for (a) $f = 25$ Hz and (b) $f = 40$ Hz.

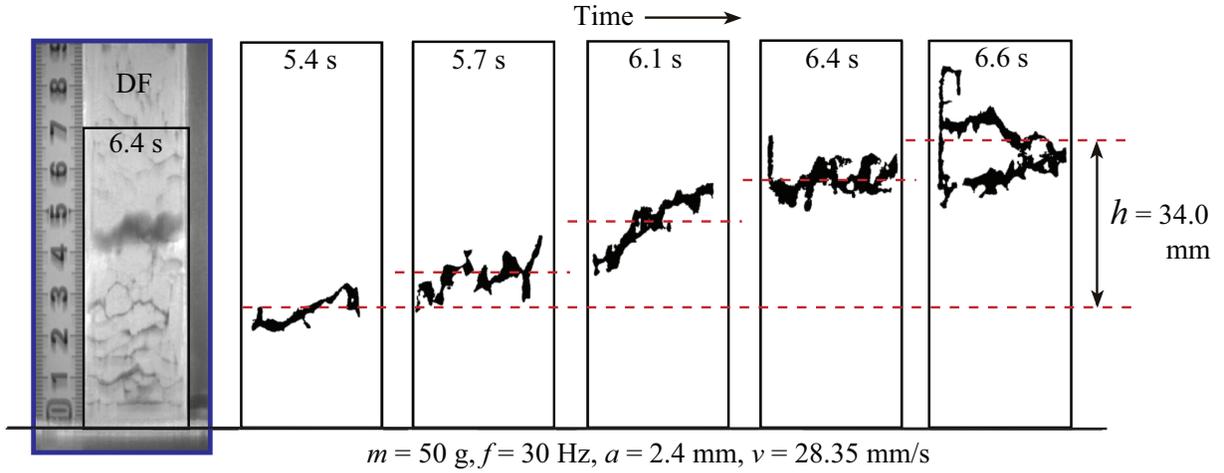

$m = 50$ g, $f = 30$ Hz, $a = 2.4$ mm, $v = 28.35$ mm/s

FIG. 5. Decompaction-wave propagation in DF regime. ImageJ [22] was used to track cracks in the bulk. The red dashed lines represent mean locations of the cracks. The standard deviation about mean of the data is defined here as a crack width $w$.

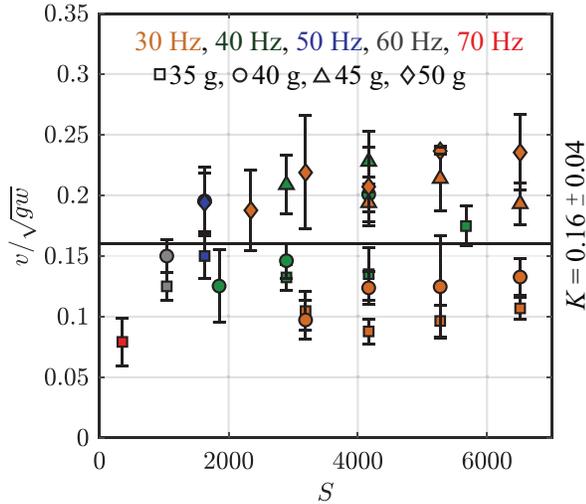

FIG. 6. The relation among $v/\sqrt{gw}$ and $S$ at RH = 40 %. $K = 0.16 \pm 0.04$ is obtained by averaging across $m$ and $f$.

### B. Cohesive strength

The relation of Eq. (1) suggests that the decompaction-wave propagation speed is solely governed by gravity, $v = K\sqrt{gw}$. The specific value of $K$ may depend on the experimental details such as geometrical setup, grain's size $d$, powder bed thickness $H$, and mass $m$. As mentioned above, the observed propagation actually consists of successive openings of the similar scale cracks from bottom to top (see the related video in SM [24]). To achieve this pseudo propagation, crack width and the interval between successive cracks should have the similar length scale $\sim w$. The macroscopic length scale is chosen for building a simple continuum-like model, where the cohesive strength $Y$ should be determined by equating it with hydrostatic pressure, $w\nu\rho g$. Thus, in DF phase, using an averaged crack width $w = 3.66$ mm from our experiments, the cohesive strength is estimated as $Y \simeq w\nu\rho g \simeq 23.7$ Pa. This estimate obtained from
5

4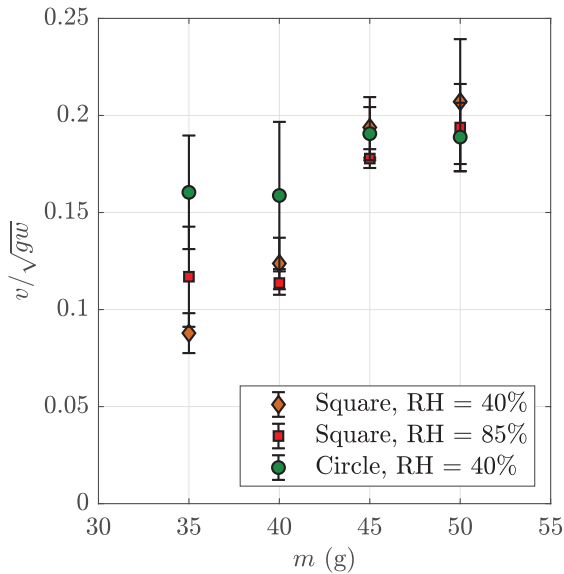

FIG. 7. Comparing $v/\sqrt{gw}$ for different relative humidities and container types. The base is vibrated at $f = 30\,\mathrm{Hz}$, and at $a = 2.4$ mm.

our phenomenological model is consistent with the experimental observations of Irie *et al.* [32] and Schweiger *et al.* [33], where strength $Y$ is estimated as 20, 32, or 89 Pa for powders in typical diameter of 5, 43, or 15 $\mu$m, respectively. Although the experimental conditions of these studies are different from the current setup, the order of obtained $Y$ values are consistent. Moreover, our estimated cohesive strength $Y$ is also supported by other available estimates [34–37]. Due to this consistency, the value of $K$ estimated in this study is close to unity. Therefore, we think that the effective decompaction-wave propagation can be observed because the hydrostatic pressure of the crack and material strength is balanced. Then, the propagation speed is governed by free fall as $\sqrt{gw}$.

In Fig. 7, RH and container shape dependence of $v/\sqrt{gw}$ is presented. As discussed earlier, the RH has negligible effect on fracturing boundary. Confirming the same, Fig. 7 shows small effect of RH on wave-velocity. When the powder container is vibrated at 30 Hz, the average width $w$ for RH = 40 % and 85 % is found to be similar. Thus, the corresponding change in cohesive strength $Y$ is also negligible. Although the $m$-dependent increase of $v/\sqrt{gw}$ can be confirmed for square shaped containers in Fig. 7, its variation range is comparable with the error of $K$ (scattering in Fig. 6). In addition, specific experiments performed with the circular shaped containers show no such clear dependence. It is difficult to conclude $m$ dependence based on the current experimental result. The $m$ dependence of $v/\sqrt{gw}$ should be evaluated by more accurate experiments.

Based on this stress-balance consideration, mechanics governing the phase boundary can also be discussed. To break apart the powder bed to form the crack, dynamic pressure by vibration $\nu\rho(a\omega)^2$ should be balanced with the strength $Y$. From this balance, the phase boundary between CS and SF (or DF) can be parameterized by $Y/\nu\rho gw = (a\omega)^2/gw = Sd/w$. Validity of such a characterization of the phase boundary will be discussed elsewhere in near future based on the systematic experiments. However, details on the crack initiation and propagation in the consolidated powder are not fully understood yet. The role of wall roughness, and the possibility of powder's stick-slip phenomenon with the wall is a topic for future research.

### C. Comparison with non-cohesive powders

Finally, we briefly comment on the behavior of vibrated powders consist of different size and density of glass beads. We performed some additional experiments using powders consisting of coarser grains that belong to group A/C on Geldart's chart [1]. For coarser grain powders, we observed convection regime (CF) instead of static fractures (SF) on the phase diagram. At higher vibrational strengths, for powders of grain size 18 $\mu$m, which lie at the boundary of group A and C as per Geldart's chart, we observed decompaction-wave propagation (DF phase) at velocity much lower than that obtained for finer powder consisting of 5 $\mu$m grain size. The corresponding $K$ value has also reduced by one order. The complex dependence of $K$ on geometrical setup, grain diameter $d$, height $H$ and mass $m$ might be the reason for this difference. Most importantly, the non-dimensional wave-velocity is found to be independent of vibrational strength $S$ (see the Appendix E for more details).

For powders belongs to group A, we observed rising bubbling layer phenomenon (see video in SM [24]). This phenomenon was first reported by Pak *et al.* [21], who used powders of grain size 200 $\mu$m. The reported bubbling effect, defined as upward moving voids, reminiscent of bubbles in fluidized bed, was observed to occur when $\Gamma$ exceeds a critical value. As the bubbles are created at the base and move upwards, their mean size grows. However, the rate of bubbling was observed to be less sensitive to change in $\Gamma$. The mechanism of forming voids/bubbles in coarse granular media is governed by the expansion of compressed air trapped in the material that occurs while detaching from the shaker base [21]. This was later verified using numerical simulation [38, 39]. When compared, in cohesive powders, we do not see any detachment. Moreover, the coarse and fine cohesive grain powders deform/fracture in ductile and brittle manner, respectively. As a consequence, cohesive powders show horizontal crack formations as the cohesive bonds among the grains break.

In summary, the results confirm universality of wave propagation in fine powders, and its independence on $S$ especially in cohesive powders. See Appendix E for more details and SM for corresponding videos [24].

## V. CONCLUSION

We conclude that the vibration of fine cohesive (group C) powders causes three distinct phases: consolidated (CS), static fracture (SF), and dynamic fracture (DF). We observed powder consolidation even at large amplitude (and low mass) regime. When the mass and amplitude exceed the thresholds, brittle fracturing was induced. In addition, we found decompaction-wave propagation in DF phase. This brittle nature of the vibrated cohesive powder is quite different from the fluidization behaviors observed in the vibrated coarse (group D) granular system. In fine cohesive powders, the observed effective decompaction-wave propagation speed is independent of vibration strength and is governed by the gravity and cohesive strength.

## VI. ACKNOWLEDGEMENTS

This work was supported by the JSPS KAKENHI, Grant No. 18H03679.

## VII. APPENDIX

### A. Real time videos of vibrated powder bed

As mentioned in the paper, the vibrated powder bed exhibits three distinct phases depending on the vibration conditions: consolidation (CS), static fracture (SF), and dynamic fracture (DF). Here, we supplement [24] following three videos of CS, SF, and DF states, which correspond to Fig. 2(c), (d), and (e), respectively: (i) CS state: CS.avi in SM shows consolidated powder when mild vibrations are employed, (ii) SF state: cracksSF.avi in SM shows formation of horizontal cracks when the vibrations are strengthened, and (iii) DF state: dwaveDF.avi in SM shows decompaction-wave propagation in the bulk when the powder is strongly vibrated. We also supplement videos [24] of vibrated powders with mean grain size $d$ of 18 $\mu$m, 50 $\mu$m, and 120 $\mu$m, respectively.

We recorded the experimental videos using cameras that capture images at 90 fps with spatial resolution 0.058 mm/pixel and 2048 × 2048 image size.

### B. Phases of vibrated powder bed

The three phases: CS, SF and DF, are differentiated based on the deformation of the void/crack. The cohesive powders are porous and the voids of various sizes and shapes can always be seen. The CS state is acquired when the voids of various sizes are no longer deforming. In SF state, we clearly observe these voids deforming into elongated cracks and fracturing the bulk. There is no quantitative criterion required to detect SF state, as the fracture is clearly visible. We qualitatively differentiate the CS and SF states. The SF state is assumed when the top free surface is fractured (see TVCStoSF.avi in SM to observe top surface fracture) and the deformed cracks start appearing in the bulk. However, for differentiating the DF state we need to employ a quantitative criterion. In DF phase, the cracks grow wider and open in a consecutive order from bottom to the top. We track the periodic upward motion of the widest crack of dimension particularly greater than 10 mm$^2$. Whereas, voids/cracks in CS and SF states rarely exceed size of 10 mm$^2$. Namely, 10 mm$^2$ is the quantitative criterion we employ to identify DF state.

Here we include some images of phase-wise transitions to clearly differentiate the three phases. We capture images showing the position of the crack versus time to illustrate the transition within CS-SF/DF states. We observed no deformation in CS regime (see Fig. 8). The small voids are steady and not deforming/moving. As shown in Fig. 9, with increased vibrational strength, we observe small cracks slowly deforming into larger cracks. These cracks are usually horizontally oriented. However, we do not see a crack wave yet. This phase is then called as static fracture due to the cracked powder bulk (see Fig. 9).

As shown in Fig. 10, we observed the transition from SF to DF, when the vibrational strength is further increased. The widened crack is the only characteristic of this transition or the boundary of SF/DF states. In general, the wide crack has area greater than 10 mm$^2$. Thus, the DF is mainly characterized by such wide crack formations. Most of the times, the crack nearest to the base starts to get widened first. After reaching the threshold subsequent consecutive cracks starts widening in upward direction. The wave velocity is not always constant through out the height. The cracks more than 10 mm$^2$ are then tracked separately in DF regime to find the wave-velocity (see Fig. 3).

### C. Formation of agglomerates

Group C powder shows cohesive characteristics forming agglomerates in which the inter-particle cohesive force dominates over gravitational force. Thus, agglomerate formation is common especially in cohesive powders when vibrated. As mentioned in this paper, the local fractures occurring at the top free surface allow the expansion of the powder bed, and thus, also result in the formation of small sized agglomerates. Throughout the experiments, we observed continuous formation and fragmentation of agglomerates largely on the top free surface. Figure 11 shows the agglomerates over consolidated powder (CS state). We limit our discussion about agglomerate formation as it corresponds to < 10 % of total mass of the powder. Although the agglomerate formation and fragmentation process is an interesting problem, we do not discuss its details in this paper.

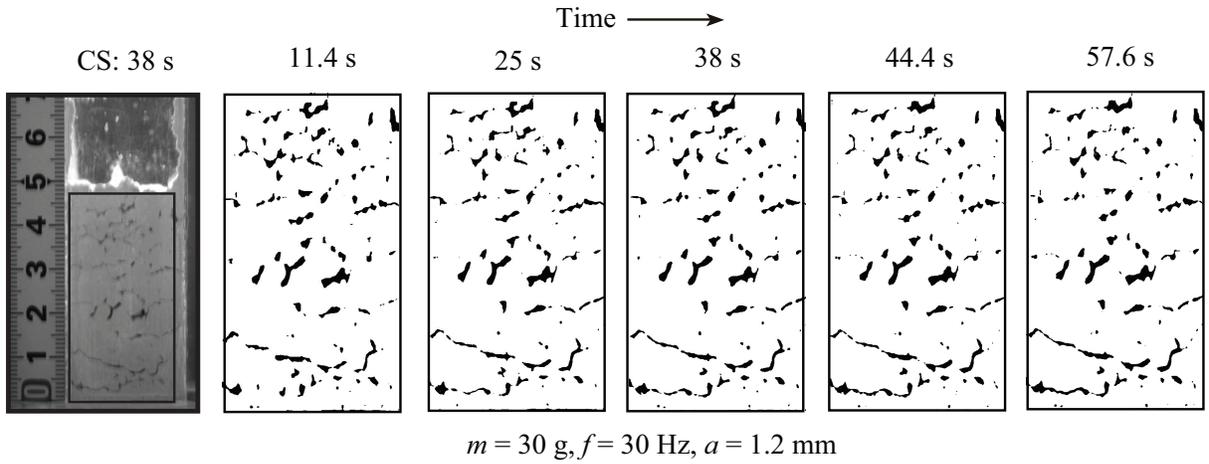

FIG. 8. Consolidated (CS) phase. ImageJ [22] is used to observe deformations in the bulk. During CS phase, we do not observe any deforming voids. The experimental conditions are $m = 30\,\text{g}$, $f = 30\,\text{Hz}$, and $a = 1.2\,\text{mm}$.

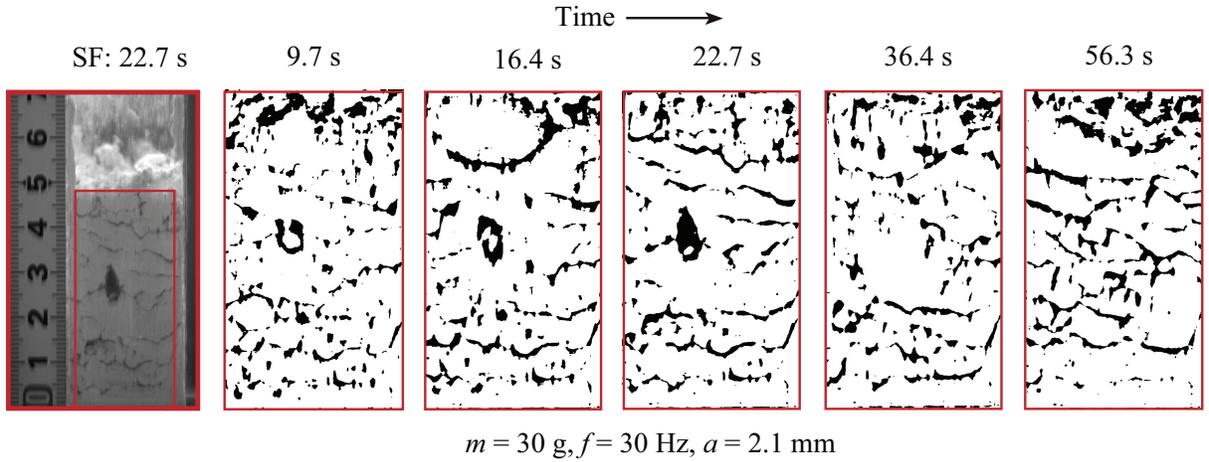

FIG. 9. Static fracturing (SF) phase. ImageJ [22] is used to observe the deformations in the bulk. The cracks/voids present in the bulk keep deforming to form elongated horizontal cracks. The experimental conditions are $m = 30\,\text{g}$, $f = 30\,\text{Hz}$, and $a = 2.1\,\text{mm}$.

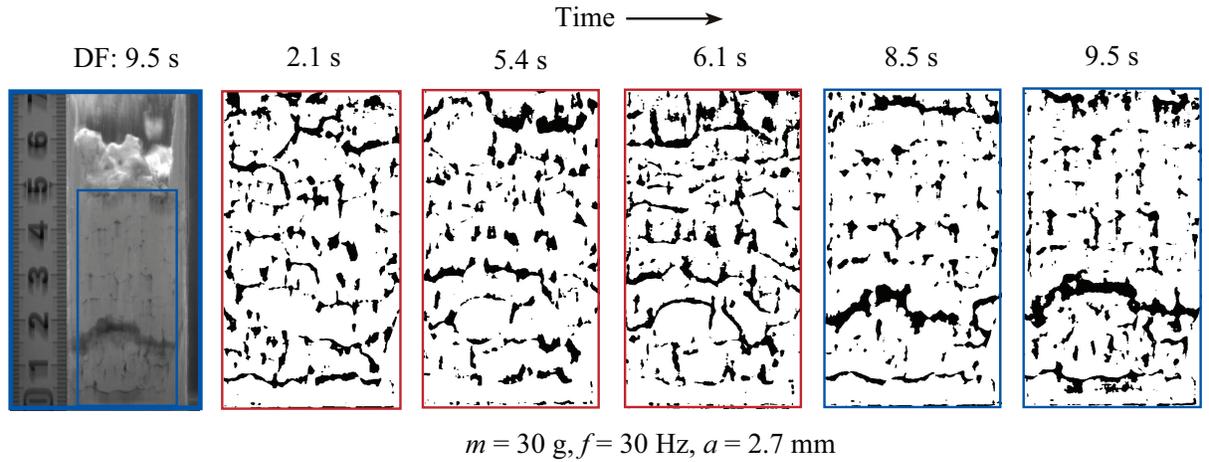

FIG. 10. Transition from static (SF) to dynamic fracturing (DF). ImageJ [22] is used to observe the deformations of widening cracks. The cracks formed during the SF phase widen when the vibrational strength is increased. The is the onset of dynamic phase. This experimental conditions are $m = 30\,\text{g}$, $f = 30\,\text{Hz}$, and $a = 2.7\,\text{mm}$.

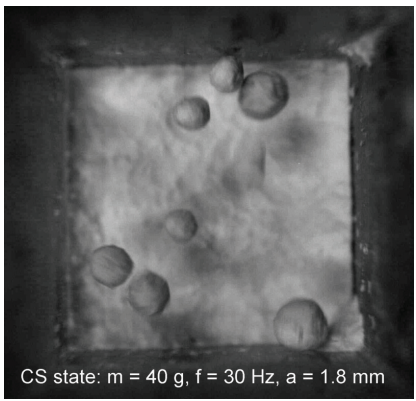

FIG. 11. Agglomerates on the top free surfaces of the consolidated powder bed. This image corresponds to the top view of the Fig. 2(c) in this paper.

Here, we supplement videos (top views) [24] showing agglomerate formation on the top free surfaces. Following videos correspond to the CS, SF and DF states, where the powder is subjected to mild and strong vibrations, respectively: (i) TVaggCS.avi in SM corresponds to the Fig. 2(c) (CS state), (ii) TVaggSF.avi in SM corresponds to the Fig. 2(d) (SF state), and (iii) TVaggDF.avi in SM corresponds to the Fig. 2(e) (DF state). In addition, a failure of top surface can be seen in TVCStoSF.avi in SM that signifies the transition from CS to SF/DF.

### D. Role of non-dimensional acceleration $\Gamma$

The relatively weak or mild vibrations can be characterized using a non-dimensional acceleration $\Gamma = a\omega^2/g$, where $a$, $\omega$, and $g$ are vibrational amplitude, angular frequency, and gravitational acceleration, respectively. Whereas, shaking strength $S = a^2\omega^2/gd$, where $d$ is the grain diameter, is better suited as a scaling parameter for strongly vibrated states. In this paper, we employed $S$ to characterize the vibration strength as the vibration amplitude is much greater than the grain size. Here, as shown in Fig. 12, any clear correlation cannot be confirmed even if we use $\Gamma$ as a scaling parameter.

### E. Coarse grain powders (Group A/C)

In order to understand the effect of grain size and density on the phases of vibrated powders, we performed some additional experiments using three different powders: (i) glass beads of the mean diameter size $d = 18$ μm and true density $\rho = 2600$ kg/m$^3$, (ii) glass beads of $d = 50$ μm and $\rho = 2600$ kg/m$^3$, and (iii) hollow low-density glass beads of $d = 120$ μm and $\rho = 900$ kg/m$^3$.

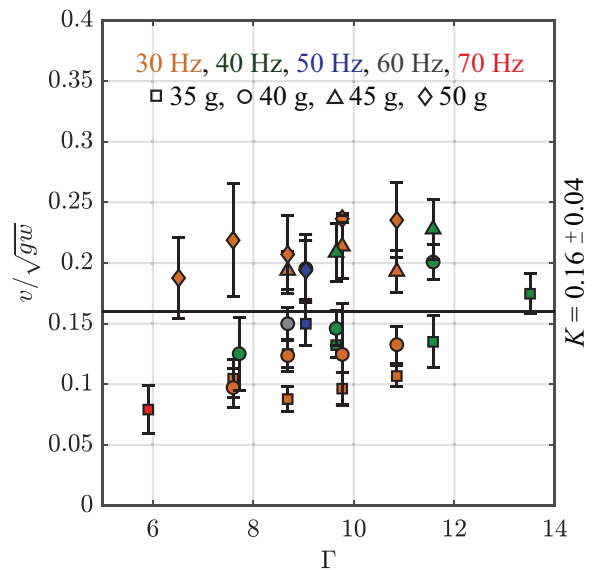

FIG. 12. The relation among $v/\sqrt{gw}$ and $\Gamma$ at RH = 40 %. $K = 0.16 \pm 0.04$ is obtained by averaging all the data.

We use Geldart's classification [1] to categorize the powders. Geldart's classification is based on experiments on dry grains fluidized with air under normal ambient conditions. The fluidized behavior of powders was found to be dependent on grain size and density. Base on the empirical observations, Geldart proposed a classification of powders as A (easily fluidized), B (causes bubbling, small bed expansion, readily fluidized), C (difficult to fluidized, fine, cohesive grains, lift as a plug, forms channels when aerated), and D (coarse grains, forms stable spouted beds) groups. Later, the empirical criteria for fluidization behavior was formulated [27], where Geldart suggested the ratio of aerated to the tapped bulk density of powder, called as Hausner ratio. Due to the contrasting behavior, a particular interest was shown in differentiating group A and C powders, which share the boundary [4, 9, 27, 40]. Roughly, the cohesive powders are known to produce high Hausner ratio (over 1.4) belongs to group C and powders producing the ratio under 1.25 belongs to group A [27]. As per the Geldart's chart based on size and density of grains, the first powder (i) we use lies at boundary between Group A and Group C, while the latter two can be classified as Group A powders. Whereas the powder we focused on ($d = 5$ μm and $\rho = 2600$ kg/m$^3$) in this paper belongs to pure group C.

The powder with 18 μm as mean diameter size that lies on the boundary of group A and C, shows mixed behavior. At low vibration strengths, small aggregates are formed on the free surface, which is one of the characteristics of cohesive powders. However, as the vibrational strength is increased, the convection was seen as in non-cohesive powders. As a consequence, we do not observe SF regime or the crack formations. As shown in Fig. 13, we observed consolidation(CS) phase, convection failure phase (CF: represented by vertical green dashed lines on phase diagram) and dynamic fracture (DF) phase. Despite of showing mixed behavior, at higher vibrational strengths i.e in DF phase, we observe the crack wave emerging from the base and moving upwards at considerably slow velocity (see dwaveDF−18.avi in SM [24]). The new wave forms as the previous wave reaches the top. More importantly, the wave-velocity is also found to be independent of vibrational strength as shown in Fig. 14(a). The measured average non-dimensional velocity was found to be $K = 0.026$ units for vibrational frequency $f = 30$ Hz. This $K$ value, about which the velocity data scatter, is much lower than that observed for powder consisting of 5 $\mu$m diameters, where $K$ was found to be 0.16.

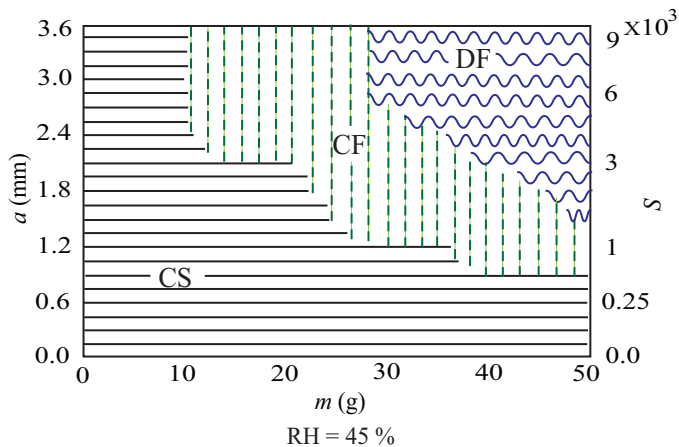

FIG. 13. The Phase diagram for a powder with mean diameter size $d = 18$ $\mu$m at $f = 30$ Hz and RH = 45%. The regime CF represents convective fracture (shown by vertical green dashed lines).

The group A powders (ii) and (iii) fluidize easily as per their characteristic, and show convection regime instead of SF regime observed in group C powders. Even at high amplitudes, we do not observe any cracks or wave propagation for $f = 30$ Hz. However, the collection of bubbles was seen near the base at $f = 40$ Hz that lifts the bed of mass exceeding 50 g. Instead of wide crack wave observed in group C powders, here the layer of bubbles is observed to be traveling upwards (see dwaveDF−50.avi in SM [24]). Such phenomenon was observed predominantly at $f = 50$ Hz for both powders (ii) and (iii) for $m \geq 50$ g. This rising layer of bubbles resembles with phenomenon reported by Pak *et al.* [21]. However, the physics behind the formation of bubble layer and its propagation is very different. The bubbling phenomenon was related to the expansion of compressed air trapped in the material and observed to occur when a critical non-dimensional acceleration $\Gamma$ is reached. This phenomenon was later confirmed by Zamankhan *et al.* [38] and Boardbar *et al.* [39] in their numerical simulations. For low-density group A powder (iii), only qualitative observations are available due to the opaque nature of light hollow spheres (see dwaveDF−120.avi in SM [24]). With limited data available of powder (ii), the dependence of non-dimensional bubble wave velocity $v$ on $S$ is not clearly established (see Fig. 14(b)). The average of $v$ is found to be scattered around $K = 0.044$.

This bubbling phenomenon observed in group A powders is different from the wave propagation phenomenon observed in our study of fine cohesive group C powders. Here we list down the following differences: (a) At low vibrational strengths, convection is observed in group A powders, while consolidation is observed in group C powders; (b) at higher vibrational strengths, voids, reminiscent of bubbles, are formed near the shaking base in group A powders, while cracks are formed throughout the cohesive consolidated bulk of group C powders; (c) with increased vibrations, a layer of collection of bubbles in group A powders travels from the base to the top, while multiple cracks from bottom to the top open in an successive order in group C powder column; (d) the group A powder grains attain free flight when detached from the vibrated base, while such detachment does not happen with cohesive grains, which do not attain a free flight at the base; and (e) the size of bubbles/voids grow as those travel upwards in group A powders, while we do not observe any increase in size of open crack in group C powders. Thus, we believe that bubbling phenomenon in group A powder is a ductile failure driven by convection, where coarse grains attain free flight. In contrast, the crack wave-propagation in cohesive, group C powders is brittle failure governed by the successive free falling of powder piles. However, to understand the mechanism of such brittle fractures in cohesive powders, the role of air and the influence of air properties must be investigated in future.

The phenomenon of wave propagation that we report in this paper is observed for a narrow band of fine powders (5 $\mu$m and 18 $\mu$m). We found that for both, cohesive powders represented by group C, and powder (i) that shows features of group A/C, the wave-velocity is independent of vibrational strength $S$. We also observe similar bubble wave-propagation for powders with non-cohesive grains (group A). However, due to the experimental limitations the dependence of bubble wave-velocity on $S$ has not been fully explored yet. At least, strong $S$ dependence could not be observed. The value of $K$ obtained for new experiments is an order of magnitude smaller than the previous experiments ($d = 5$ $\mu$m). We think the property of fine powders to go through the brittle fracturing mode from single coherent (CS) phase to dynamic (DF) phase is an important difference. In coarser grains, we observed convection instead of fracturing or cracking. We believe these small $K$ values might mean that bubble rising in coarse grains is not governed solely by gravity as estimated in cohesive powders. Thus, $v$ may have dependence on grain size $d$, vibrational angular frequency $\omega$, and the bed height $H$. In such case, the wave propagation speed may be governed as $v = K(d\omega)^{2\alpha}(gw)^{1/2-\alpha}$ in terms of dimensional consistency. Here, $K$ is a function of $H$, and $\alpha$ is an arbitrary parameter less than a half. The specific form of such relation can only be confirmed in future by performing experiments for various geometric factors and for range of non-cohesive coarse grain powders. For the current study we focus on fine cohesive grains only.

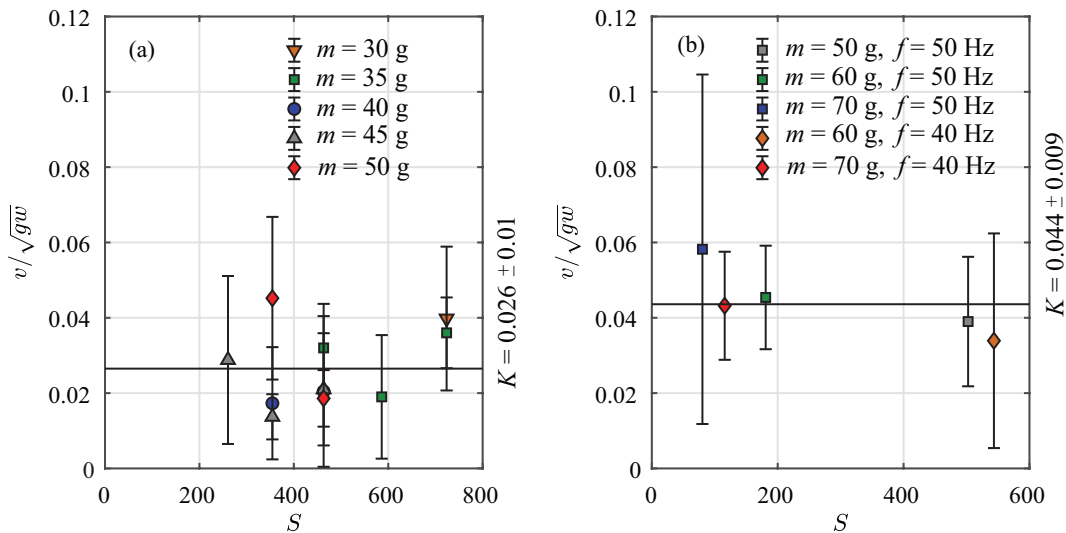

FIG. 14. The relation among $v/\sqrt{gw}$ and $S$ for powders: (a) $d = 18$ μm at 30 Hz, and (b) $d = 50$ μm at two different frequencies.